\documentclass[prd,eqsecnum,aps,twocolumn,nofootinbib]{revtex4-2}

\usepackage{amsfonts}
\usepackage{amsmath}
\usepackage{amssymb}
\usepackage{amsthm}
\usepackage{bm}
\usepackage{dcolumn}
\usepackage{epsfig}
\usepackage{graphicx}
\usepackage{graphics}
\usepackage[latin1]{inputenc}
\usepackage{latexsym}

\begin{document}

\title{Parameterizing black hole orbits for adiabatic inspiral}

\author{Scott A.\ Hughes}
\address{Department of Physics and MIT Kavli Institute, Massachusetts Institute of Technology, Cambridge, MA 02139}

\begin{abstract}
Adiabatic binary inspiral in the small mass ratio limit treats the small body as moving along a geodesic of a large Kerr black hole, with the geodesic slowly evolving due to radiative backreaction.  Up to initial conditions, geodesics are typically parameterized in two ways: using the integrals of motion energy $E$, axial angular momentum $L_z$, and Carter constant $Q$; or, using orbit geometry parameters semi-latus rectum $p$, eccentricity $e$, and (cosine of ) inclination $x_I \equiv \cos I$.  The community has long known how to compute orbit integrals as functions of the orbit geometry parameters, i.e., as functions expressing $E(p, e, x_I)$, and likewise for $L_z$ and $Q$.  Mappings in the other direction --- functions $p(E, L_z, Q)$, and likewise for $e$ and $x_I$ --- have not yet been developed in general.  In this note, we develop generic mappings from ($E$, $L_z$, $Q$) to ($p$, $e$, $x_I$).  The mappings are particularly simple for equatorial orbits ($Q = 0$, $x_I = \pm1$), and can be evaluated efficiently for generic cases.  These results make it possible to more accurately compute adiabatic inspirals by eliminating the need to use a Jacobian which becomes singular as inspiral approaches the last stable orbit.
\end{abstract}

\maketitle

\section{Introduction: Kerr black hole orbits and adiabatic inspiral}

The leading order model for computing the inspiral of small bodies into large black holes is known as ``adiabatic'' inspiral.  Adiabatic inspirals use orbit-averaged radiation reaction to approximate the inspiral of a small-mass-ratio binary as the slow evolution of a black hole's geodesic orbits.  The framework for computing adiabatic inspirals and their associated gravitational waveforms is well developed \cite{hughes2021}, and forms the basis of the Fast EMRI Waveform, or ``FEW,'' program \cite{FEW1, FEW2, FEW3}.  FEW is likely to play an important role in studies assessing the ability of the low-frequency gravitational-wave detector LISA \cite{lisa} to measure ``extreme mass-ratio inspirals,'' or EMRIs, small mass-ratio binaries formed by the capture of stellar-mass (roughly $10-100\,M_\odot$) compact objects onto strong-field orbits of massive black holes (roughly $10^5-10^7\,M_\odot$) in the cores of galaxies.  Adiabatic models also serve as a foundation for understanding post-adiabatic effects.  Many important post-adiabatic effects can likely be incorporated into waveform models with slight augmentations to the adiabatic waveform framework already developed.  It is thus of great interest to develop accurate tools and data for computing precise adiabatic inspirals and waveforms, covering the astrophysical parameter space expected for EMRI events.

Bound geodesic orbits of Kerr black holes are typically described using the orbit integrals energy $E$, axial angular momentum $L_z$, and Carter constant $Q$; or, by a set of parameters which characterize an orbit's geometry, ($p$, $e$, $x_I$).  The parameters $p$ and $e$ describe the range of radial motion over which the orbit moves: in Boyer-Lindquist coordinates (which we use in this analysis), an orbit oscillates between periapsis and apoapsis radii
\begin{equation}
r_{\rm a} = \frac{p}{1 - e}\;,\quad r_{\rm p} = \frac{p}{1 + e}\;.
\label{eq:radialrange}
\end{equation}
The parameter $x_I \equiv \cos I$ determines an orbit's inclination to the black hole's equatorial plane: the orbit oscillates in polar angle $\theta$ between $\theta_{\rm min}$ and $\pi - \theta_{\rm min}$, where $I$ and $\theta_{\rm min}$ are related by
\begin{equation}
I = \pi/2 - {\rm sgn}(L_z)\theta_{\rm min}\;,
\label{eq:incldef}
\end{equation}
which in turn means that
\begin{equation}
x_I \equiv {\rm sgn}(L_z)\sin(\theta_{\rm min})\;.
\label{eq:xIdef}
\end{equation}
The angle $I$ is $0$ for prograde equatorial orbits, $180^\circ$ for retrograde equatorial, and smoothly varies between these extremes.  Note that the sign of $x_I$ usefully encodes the prograde or retrograde character of black hole orbits.

Data describing backreaction and inspiral waveforms are typically parameterized using ($p$, $e$, $x_I$), since these parameters determine the properties of the source term to the equations of black hole perturbation theory.  To make an inspiral, we need to compute the rates of change of these parameters, ($dp/dt$, $de/dt$, $dx_I/dt$).  Adiabatic backreaction calculations directly determine the rates at which the orbit integrals change, ($dE/dt$, $dL_z/dt$, $dQ/dt$).  It is simple to write down a Jacobian which yields ($dp/dt$, $de/dt$, $dx_I/dt$) given ($dE/dt$, $dL_z/dt$, $dQ/dt$).  As a matter of principle, computing adiabatic inspiral is straightforward once the rates of change ($dE/dt$, $dL_z/dt$, $dQ/dt$) are known.

Unfortunately, that Jacobian becomes singular as we approach the last stable orbit or LSO.  This behavior can be seen particularly clearly by examining the simplest limit, circular inspiral into a Schwarzschild black hole.  Such an orbit is characterized by its radius $r_{\rm o}$ (which is equivalent to $p$ when $e = 0$), which in turn determines orbital energy per unit mass $E$:
\begin{equation}
E(r_{\rm o}) = \frac{1 - 2M/r_{\rm o}}{\sqrt{1 - 3M/r_{\rm o}}}\;.
\label{eq:Eschw}
\end{equation}
(We use units with $G = 1$, $c = 1$.)  Suppose that we have computed $dE/dt$ at many values of $r_{\rm o}$.  Using this dataset and this functional form, we find $dr_{\rm o}/dt$:
\begin{equation}
\frac{dr_{\rm o}}{dt} = \frac{dE/dt}{dE/dr_{\rm o}} = \frac{2(r_{\rm o}(r_{\rm o} - 3M))^{3/2}}{M(r_{\rm o} - 6M)}\frac{dE}{dt}\;.
\label{eq:Jacobian_circschw}
\end{equation}
Note the singularity as $r_{\rm o} \to 6M$.  Here, geodesic orbits are marginally stable, so a small push (small $dE/dt$) produces a large response (large $dr_{\rm o}/dt$), making $dr_{\rm o}/dt$ stiff as we approach $r_{\rm o} = 6M$.  Suppose that we have stored $dr_{\rm o}/dt$ on a grid of orbits that covers the domain from close to the LSO $r_{\rm o} = 6M$ to some large radius.  These data will be difficult to interpolate on this grid, and we will lose numerical accuracy as we approach $r_{\rm o} = 6M$.  We can in principle compensate by using a very dense grid near the LSO and by taking small step sizes in this region.  However, these fixes only postpone the difficulty, which fundamentally reflects the change from stable to unstable orbits at this radius.

A better approach is to work with data from which the pathology has been removed.  In Eq.\ (\ref{eq:Jacobian_circschw}), the singularity is simple enough that it can be removed analytically: rather than storing $dr_{\rm o}/dt$ on our grid, we store $\dot y(r_{\rm o}) \equiv (r_{\rm o} - 6M)dr_{\rm o}/dt$.  Data for $\dot y(r_{\rm o})$ are smooth all the way to the LSO; they can be interpolated accurately over the entire domain of inspiral and used to compute $dr_{\rm o}/dt$ accurately.

Unfortunately, the singularity is more complicated in the general case than what we see in Eq.\ (\ref{eq:Jacobian_circschw}), and is not as amenable to such a simple ``repair.''  A more robust (and arguably more elegant) approach to curing the fundamental pathology is to parameterize using a quantity that is never singular.  For circular and equatorial orbits, energy is such a parameter.  Though the rate of change $dE/dt$ grows as one moves into the strong field, it does so in a way that is smooth and well behaved.  Beginning at some initial orbit, it is straightforward to integrate $dE/dt$ to construct $E(t)$, the orbital energy along the inspiral worldline's sequence of orbits.  If one needs orbital radius along the inspiral (say because backreaction and waveform data have been stored on a grid parameterized with $r_{\rm o}$), it is simple to invert Eq.\ (\ref{eq:Eschw}):
\begin{equation}
r_{\rm o}(E) = \frac{8M}{(4 - 3E^2 - E^2\sqrt{9E^2 - 8})}\;.
\label{eq:Eschwinv}
\end{equation}
As energy varies over $2\sqrt{2}/3 \le E < 1$, Eq.\ (\ref{eq:Eschwinv}) yields the radii of all stable\footnote{A second solution, with opposite sign for the square root in the denominator, yields unstable orbit radii in $4M < r_{\rm o} \le 6M$.} circular orbits in $6M \le r_{\rm o} < \infty$.

Although we have focused in this introductory discussion on circular orbits of Schwarzschild black holes, this core pathology holds for all black hole orbits: the Jacobian between ($dE/dt$, $dL_z/dt$, $dQ/dt$) and ($dp/dt$, $de/dt$, $dx_I/dt$) is singular as we approach the LSO, so using ($dp/dt$, $de/dt$, $dx_I/dt$) will almost certainly result in poor numerical accuracy when computing adiabatic inspiral.  Switching the parameterization from ($p$, $e$, $x_I$) to ($E$, $L_z$, $Q$) cures this pathology in the general case.

A challenge to implementing this cure is that the ``inverse'' mapping, from integrals of motion to orbital geometry parameters, has not been developed for the general case.  In this paper, we develop and present this inverse mapping for general Kerr geodesics.  We begin by briefly reviewing the relevant properties of Kerr geodesics in Sec.\ \ref{sec:geods}.  In Sec.\ \ref{sec:equatorial}, we develop the mapping from ($E$, $L_z$) to ($p$, $e$) for equatorial orbits (which have $Q = 0$ and $x_I = \pm 1$), for which the mapping is particularly simple.  We present the solution for the general case in Sec.\ \ref{sec:generic}.  In Sec.\ \ref{sec:compare}, we compare inspirals that use these two parameterizations.  Concluding discussion, including plans to use these results to extend the parameter coverage of FEW, is given in Sec.\ \ref{sec:conclusions}.

\section{Key properties of Kerr black hole bound geodesic orbits}
\label{sec:geods}

In Boyer-Lindquist coordinates, the equations governing orbits of a small body of mass $\mu \ll M$ about a Kerr black hole with mass $M$ and spin parameter $a = |{\bf S}|/M$ (where ${\bf S}$ is the hole's spin angular momentum) are
\begin{eqnarray}
\left(\frac{dr}{d\lambda}\right)^2 &=& [E(r^2 + a^2) - aL_z]^2
\nonumber\\
&& - \Delta[r^2 + (L_z - aE)^2 + Q] \equiv R(r)\;,
\label{eq:rdot}\\
\left(\frac{d\theta}{d\lambda}\right)^2 &=& Q - \cot^2\theta L_z^2 - a^2\cos^2\theta(1 - E^2)
\nonumber\\
&\equiv& \Theta(\theta)\;,
\label{eq:thdot}\\
\frac{d\phi}{d\lambda} &=& \csc^2\theta L_z + \frac{2Mr a E}{\Delta} - \frac{a^2 L_z}{\Delta}\;,
\label{eq:phdot}\\
\frac{dt}{d\lambda} &=& E\left[\frac{(r^2 + a^2)^2}{\Delta} - a^2\sin^2\theta\right] - \frac{2Mra L_z}{\Delta}\;.
\nonumber\\
\label{eq:tdot}
\end{eqnarray}
The time parameter $\lambda$ we use here is known as {\it Mino time}; an interval of Mino time $d\lambda$ is related to an interval of proper time along the orbit $d\tau$ by $d\lambda = d\tau/\Sigma$, where $\Sigma = r^2 + a^2\cos^2\theta$.  We have introduced $\Delta = r^2 - 2Mr + a^2$.  The quantities $E$, $L_z$, and $Q$ are the orbit's energy (per unit $\mu$), axial angular momentum (per unit $\mu$), and Carter constant (per unit $\mu^2$).  These quantities are conserved along any geodesic; choosing them specifies an orbit, up to initial conditions.  Writing $d/d\lambda = \Sigma\, d/d\tau$ puts these equations into more familiar forms typically found in textbooks, such as Eqs.\ (33.32a--d) of Ref.\ \cite{mtw}.

The functions $R(r)$ and $\Theta(\theta)$ determine an orbit's properties.  In particular, some of the zeros of these functions tell us about an orbit's turning points, and some of the zeros of $R(r)$ encode whether an orbit is stable or not.  Consider the polar motion first.  The orbit's polar velocity goes to zero at an angle $\theta_{\rm min}$ defined by
\begin{equation}
Q - \cot^2\theta_{\rm min}L_z^2 - a^2\cos^2\theta_{\rm min}(1 - E^2) = 0\;.
\label{eq:thetaturningpoint}
\end{equation}
This condition tells us that the polar motion stops and reverses direction at $\theta = \theta_{\rm min}$, which is the minimum value of $\theta$ reached by the orbit.  Thanks to reflection symmetry about the Kerr equatorial plane $\theta = \pi/2$, the polar motion likewise stops and reverses at $\theta = \theta_{\rm max} = \pi - \theta_{\rm min}$, the maximum value of $\theta$ reached by the orbit.

Rewriting Eq.\ (\ref{eq:thetaturningpoint}) in terms of the cosine of inclination defined by Eq.\ (\ref{eq:incldef}) yields a quadratic equation for $x_I^2$:
\begin{equation}
0 = a^2(1 - E^2)x_I^4 + \left(Q + L_z^2 - a^2(1 - E^2)\right)x_I^2 - L_z^2 = 0\;.
\label{eq:xIeqn}
\end{equation}
Solving this yields $x_I$ as a function of the integrals of motion.  We write this solution
\begin{widetext}
\begin{equation}
x_I(E, L_z, Q) = \frac{\sqrt{2}L_z}{\sqrt{Q + L_z^2 - a^2(1 - E^2) + \sqrt{\left(Q + L_z^2 - a^2(1 - E^2)\right)^2 + 4a^2L_z^2(1 - E^2)}}}\;.
\label{eq:xIfunc}
\end{equation}
\end{widetext}
This form is written to be well-behaved as $a \to 0$.  It is straightforward to show that Eq.\ (\ref{eq:xIfunc}) respects the equatorial limit: $x_I \to L_z/|L_z|$ as $Q \to 0$.

Inverse solutions for the other two parameters are found by examining the radial equation.  We present detailed solutions in the following two sections, beginning here with general considerations on the properties of these solutions.  We start by carefully examining $R(r)$, which we write in two different ways:
\begin{eqnarray}
R(r) &=& (E^2 - 1)r^4 + 2M r^3 +[a^2(E^2 - 1) - L_z^2 - Q]r^2
\nonumber\\
& & + 2M[Q + (aE - L_z)^2]r - a^2Q\;,
\label{eq:quartic}\\
&=& (1 - E^2)(r_1 - r)(r - r_2)(r - r_3)(r - r_4)\;.
\label{eq:Rroots}
\end{eqnarray}
Equation (\ref{eq:quartic}) is what we get when we write Eq.\ (\ref{eq:rdot}) as an explicit polynomial in $r$; Eq.\ (\ref{eq:Rroots}) rewrites this in a way that emphasizes its four roots.  These roots are ordered such that $r_1 \ge r_2 \ge r_3 > r_4$.  The root $r_4$ is inside the event horizon, and so is never reached by bound orbits.  Indeed, $r_4 = 0$ for Schwarzschild black holes ($a = 0$) and equatorial orbits ($Q = 0$).  The properties of bound orbits are thus determined by the values of $r_1$, $r_2$, and $r_3$.  Four cases are particularly important:

\begin{itemize}

\item {\it Stable eccentric orbits}: $r_1 > r_2 > r_3$.  The orbiting body's coordinate radial velocity reverses, passing through zero, when $r = r_1$ and $r = r_2$.  When the three outer roots are distinct, one can show that the coordinate acceleration $d^2r/d\lambda^2$ does not vanish at these turning points, but has a sign which insures that the motion oscillates between these points.

\item {\it Stable circular orbits}: $r_1 = r_2 > r_3$.  When the two outermost roots coincide but are distinct from the third, both the coordinate velocity $dr/d\lambda$ and the coordinate acceleration $d^2r/d\lambda^2$ vanish at $r = r_1$, so the orbiting body sits at this radius for all time.  Further analysis shows that this orbital motion is stable against small perturbations.

\item {\it Marginally stable eccentric orbits}: $r_1 > r_2 = r_3$.  In this case, the outer root at $r = r_1$ remains a turning point, but both radial coordinate velocity and coordinate acceleration vanish at $r = r_2$.  After reversing direction at $r = r_1$, a geodesic with this configuration will ``whirl'' eternally at $r = r_2$.  The inner turning point is not stable; the condition $r_2 = r_3$ defines the last or innermost stable orbit.

\item {\it Marginally stable circular orbits}: $r_1 = r_2 = r_3$.  This combines the circularity condition $r_1 = r_2$ with the marginal stability condition $r_2 = r_3$.  This defines the innermost stable circular orbit.

\end{itemize}

For eccentric orbits, it is useful to relabel the outer two roots, which define the range of the smaller body's motion: we call the outermost root the apoapsis, and the next root in the periapsis: $r_1 = r_{\rm a}$, $r_2 = r_{\rm p}$.  Using Eq.\ (\ref{eq:radialrange}), we remap these radii to the semi-latus rectum $p$ and eccentricity $e$:
\begin{equation}
p = \frac{2r_{\rm a}r_{\rm p}}{r_{\rm a} + r_{\rm p}}\;,\qquad e = \frac{r_{\rm a} - r_{\rm p}}{r_{\rm a} + r_{\rm p}}\;.
\label{eq:pe_of_rarp}
\end{equation}
For circular orbits, the coinciding outer roots are simply the orbit's radius: $r_1 = r_2 = r_{\rm o}$.  In all cases, we have closed-form expressions for $E(p, e, x_I)$, and likewise for $L_z$ and $Q$; see Refs.\ \cite{schmidt, fujitahikida, vandemeent} for explicit formulas.

\section{Converting the parameters of equatorial orbits}
\label{sec:equatorial}

Our goal is now to find expressions for the roots $r_{1,2,3,4}$ given an orbit's integrals of motion $E$, $L_z$, and $Q$ (plus the Kerr parameter $a$).  We begin with equatorial orbits, for which $Q = 0$ and $r_4 = 0$.  Equations (\ref{eq:quartic}) and (\ref{eq:Rroots}) simplify in this case to
\begin{eqnarray}
R_{\rm eq}(r) &=& r\Bigl[(E^2 - 1)r^3 + 2Mr^2 + [a^2(E^2 - 1)-L_z^2]r
\nonumber\\
& & + 2M(aE - L_z)^2\Bigr]
\label{eq:quartic_eq}
\\
&=& (1 - E^2)r\Bigl[(r_1 - r)(r - r_2)(r - r_3)\Bigr]\;.
\label{eq:Rroots_eq}
\end{eqnarray}
Taking advantage of the trivial root $r_4 = 0$, what remains is a cubic.  Closed-form solutions for cubic equations with real coefficients yielding real roots have been known since 1615 \cite{viete}; our discussion follows that presented in Ref.\ \cite{numrec}.  Dividing Eq.\ (\ref{eq:quartic_eq}) by $r(E^2 - 1)$, the polynomial whose roots we wish to find is given by
\begin{equation}
R_3(r) = r^3 + \mathcal{A}_2r^2 + \mathcal{A}_1r + \mathcal{A}_0\;,
\end{equation}
where
\begin{eqnarray}
\mathcal{A}_2 &=& \frac{2M}{(E^2 - 1)}\;,
\\
\mathcal{A}_1 &=& \frac{a^2(E^2 - 1) - L_z^2}{(E^2 - 1)}\;,
\\
\mathcal{A}_0 &=& \frac{2M(aE - L_z)^2}{(E^2 - 1)}\;.
\end{eqnarray}
Define
\begin{eqnarray}
\mathcal{Q} &=& \frac{1}{9}\left(\mathcal{A}_2^2 - 3\mathcal{A}_1\right)\;,\;
\label{eq:Q}\\
\mathcal{R} &=& \frac{1}{54}\left(2\mathcal{A}_2^3 - 9\mathcal{A}_2\mathcal{A}_1 + 27\mathcal{A}_0\right)
\label{eq:R}
\end{eqnarray}
and
\begin{equation}
\vartheta = \arccos\left(\mathcal{R}/\sqrt{\mathcal{Q}^3}\right)\;.
\label{eq:theta}
\end{equation}
Then, the roots governing equatorial Kerr black hole orbits are given by
\begin{eqnarray}
r_{\rm a} &=& -2\sqrt{\mathcal{Q}}\cos\left(\frac{\vartheta + 2\pi}{3}\right) - \frac{\mathcal{A}_2}{3}\;,
\label{eq:r_a_eq}\\
r_{\rm p} &=& -2\sqrt{\mathcal{Q}}\cos\left(\frac{\vartheta - 2\pi}{3}\right) - \frac{\mathcal{A}_2}{3}\;,
\label{eq:r_p_eq}\\
r_3 &=& -2\sqrt{\mathcal{Q}}\cos\left(\frac{\vartheta}{3}\right) - \frac{\mathcal{A}_2}{3}\;.
\label{eq:r_3_eq}
\end{eqnarray}
With $r_{\rm a}$ and $r_{\rm p}$ known, we find $p$ and $e$ using Eq.\ (\ref{eq:pe_of_rarp}).  Since originally posting this paper, we have learned that a similar solution to Eqs.\ (\ref{eq:r_a_eq})--(\ref{eq:r_3_eq}), focusing on unbound orbits of Schwarzschild black holes and based on unpublished work by van de Meent, appears in Ref.\ \cite{bl22}.

In the supplemental material accompanying this paper, we provide a {\it Mathematica} notebook which implements these formulas, as well as Eq.\ (\ref{eq:xIfunc}).  It is straightforward to verify that these results accurately yield $p(E, L_z)$, $e(E, L_z)$, and $r_3(E, L_z)$.  For example, using the {\it Black Hole Perturbation Toolkit} package {\tt KerrGeodesics} \cite{bhpt}, compute $E(p, e)$, $L_z(p, e)$, and $r_3(p, e)$ for an equatorial Kerr orbit.  Inserting the numerical results for $E$ and $L_z$ into Eqs.\ (\ref{eq:r_a_eq})--(\ref{eq:r_3_eq}) and using (\ref{eq:pe_of_rarp}) returns the original values of $p$ and $e$, and confirms $r_3$.  A {\tt C}-code implementation of these formulas (also provided with the supplemental material) likewise validates these results to machine precision.

It's worth noting that $\cos\vartheta = \mathcal{R}/\sqrt{\mathcal{Q}^3}$ has magnitude less than 1 for all stable bound geodesics, and is equal to 1 for marginally stable geodesics.  This is straightforward to show for Schwarzschild: using the closed-form solutions for $E$ and $L_z$ in this limit,
\begin{eqnarray}
E_{\rm Schw} &=& \sqrt{\frac{\left(p - 2M(1 + e)\right)\left(p - 2M(1 - e)\right)}{p\left(p - M(3 + e^2)\right)}}\;,
\nonumber\\
L_{z,{\rm Schw}} &=& \sqrt{\frac{p^2M}{p - M(3 + e^2)}}\;,
\end{eqnarray}
we find $\cos\vartheta < 1$ for all orbits outside of the Schwarzschild LSO at $(6+2e)M$, and $\cos\vartheta = 1$ exactly on the LSO.  (This calculation is done in the {\it Mathematica} notebook included in this paper's auxiliary material.)  For Kerr, this evaluation is not so clean because the expressions for $E$ and $L_z$ are more complicated, but one can nonetheless validate this behavior.  Our {\tt C}-code checks whether $\cos\vartheta \le 1$; if it is not, the code exits with a warning that the parameters provided do not correspond to bound stable orbits.  The {\it Mathematica} notebook will evaluate Eqs.\ (\ref{eq:r_a_eq})--(\ref{eq:r_3_eq}) even for parameters inside the LSO, but returns roots off the real axis in such cases.

\section{Converting the parameters of generic orbits}
\label{sec:generic}

For generic orbits, the radial function does not simplify, and we must find all four of its roots.  Closed-form solutions for the real roots of real quartic polynomials have also been known for quite some time.  The solution we use was developed by Euler \cite{euler}, though we follow the slightly modified algorithm given by Wolters \cite{wolters}.  Begin by dividing Eq.\ (\ref{eq:quartic}) by $E^2 - 1$, and writing the resulting quartic
\begin{equation}
R_4(r) = r^4 + \mathcal{A}_3 r^3 + \mathcal{A}_2r^2 + \mathcal{A}_1r + \mathcal{A}_0\;,
\end{equation}
defining the coefficients
\begin{eqnarray}
\mathcal{A}_3 &=& \frac{2M}{(E^2 - 1)}\;,
\label{eq:A3gen}\\
\mathcal{A}_2 &=& \frac{a^2(E^2 - 1) - L_z^2 - Q}{(E^2 - 1)}\;,
\label{eq:A2gen}\\
\mathcal{A}_1 &=& \frac{2M\left(Q + (a E - L_z)^2\right)}{(E^2 - 1)}\;,
\label{eq:A1gen}\\
\mathcal{A}_0 &=& -\frac{a^2Q}{(E^2 - 1)}\;.
\label{eq:A0gen}
\end{eqnarray}
Define further
\begin{eqnarray}
\mathcal{B}_2 &=& \mathcal{A}_2 - \frac{3\mathcal{A}_3^2}{8}\;,
\label{eq:B2}\\
\mathcal{B}_1 &=& \mathcal{A}_1 - \frac{\mathcal{A}_2\mathcal{A}_3}{2} + \frac{\mathcal{A}_3^3}{8}\;,
\label{eq:B1}\\
\mathcal{B}_0 &=& \mathcal{A}_0 - \frac{\mathcal{A}_1\mathcal{A}_3}{4} + \frac{\mathcal{A}_2\mathcal{A}_3^2}{16} - \frac{3\mathcal{A}_3^4}{256}\;.
\end{eqnarray}
These $\mathcal{B}_n$ coefficients appear in the {\it resolvent cubic} function,
\begin{eqnarray}
f_{\rm rc}(z) = z^3 + \mathcal{C}_2z^2 + \mathcal{C}_1z + \mathcal{C}_0\;,
\label{eq:resolvent}
\end{eqnarray}
where the coefficients introduced here are related to quantities defined above by
\begin{equation}
\mathcal{C}_2 = \frac{\mathcal{B}_2}{2}\;,\quad
\mathcal{C}_1 = \frac{\mathcal{B}_2^2}{16} - \frac{\mathcal{B}_0}{4}\;,\quad
\mathcal{C}_0 = -\frac{\mathcal{B}_1^2}{64}\;.
\label{eq:Cn}
\end{equation}
The resolvent cubic has the dimension of length to the sixth power; its three roots are real across the parameter space of bound Kerr orbits, only becoming complex if we examine orbits inside the last stable orbit.  To find these roots, first define
\begin{eqnarray}
\mathcal{Q}_{\rm rc} &=& \frac{1}{9}\left(\mathcal{C}_2^2 - 3\mathcal{C}_1\right)\;,\;
\label{eq:Qrc}\\
\mathcal{R}_{\rm rc} &=& \frac{1}{54}\left(2\mathcal{C}_2^3 - 9\mathcal{C}_2\mathcal{C}_1 + 27\mathcal{C}_0\right)\;,
\label{eq:Rrc}\\
\vartheta_{\rm rc} &=& \arccos\left(\mathcal{R}_{\rm rc}/\sqrt{\mathcal{Q}_{\rm rc}^3}\right)\;.
\end{eqnarray}
Then,
\begin{eqnarray}
z_{{\rm rc}1} &=& -2\sqrt{\mathcal{Q_{\rm rc}}}\cos\left(\frac{\vartheta_{\rm rc} + 2\pi}{3}\right) - \frac{\mathcal{C}_2}{3}\;,
\label{eq:zcrt1}\\
z_{{\rm rc}2} &=& -2\sqrt{\mathcal{Q_{\rm rc}}}\cos\left(\frac{\vartheta_{\rm rc} - 2\pi}{3}\right) - \frac{\mathcal{C}_2}{3}\;,
\label{eq:zcrt2}\\
z_{{\rm rc}3} &=& -2\sqrt{\mathcal{Q_{\rm rc}}}\cos\left(\frac{\vartheta_{\rm rc}}{3}\right) - \frac{\mathcal{C}_2}{3}\;.
\label{eq:zcrt3}
\end{eqnarray}
The roots $z_{{\rm rc}n}$, which each have the dimension of length squared, allow us to assemble the roots of the Kerr radial function $R(r)$:
\begin{widetext}
\begin{eqnarray}
r_{\rm a} &=& \frac{M}{2(1 - E^2)} + \sqrt{z_{{\rm rc}1}} + \sqrt{z_{{\rm rc}2} + z_{{\rm rc}3} - 2{\rm sgn}(\mathcal{B}_1)\sqrt{z_{{\rm rc}2}z_{{\rm rc}3}}}\;,
\label{eq:r_a_gen}\\
r_{\rm p} &=& \frac{M}{2(1 - E^2)} + \sqrt{z_{{\rm rc}1}} - \sqrt{z_{{\rm rc}2} + z_{{\rm rc}3} - 2{\rm sgn}(\mathcal{B}_1)\sqrt{z_{{\rm rc}2}z_{{\rm rc}3}}}\;,
\label{eq:r_p_gen}\\
r_{\rm 3} &=& \frac{M}{2(1 - E^2)} - \sqrt{z_{{\rm rc}1}} + \sqrt{z_{{\rm rc}2} + z_{{\rm rc}3} + 2{\rm sgn}(\mathcal{B}_1)\sqrt{z_{{\rm rc}2}z_{{\rm rc}3}}}\;,
\label{eq:r_3_gen}\\
r_{\rm 4} &=& \frac{M}{2(1 - E^2)} - \sqrt{z_{{\rm rc}1}} - \sqrt{z_{{\rm rc}2} + z_{{\rm rc}3} + 2{\rm sgn}(\mathcal{B}_1)\sqrt{z_{{\rm rc}2}z_{{\rm rc}3}}}\;.
\label{eq:r_4_gen}
\end{eqnarray}
\end{widetext}
We again emphasize that it is simple to convert ($r_{\rm a}$, $r_{\rm p}$) to ($p$, $e$) using Eq.\ (\ref{eq:pe_of_rarp}).

We have found that this method yields robust and accurate solutions across the parameter space of generic bound Kerr geodesic orbits.  During this study, we also examined Ferrari's method, which corresponds to the solution returned by the {\tt Solve[]} function of {\it Mathematica}; see also discussion in Ref.\ \cite{wolters}.  (Indeed, we learned that an inverse mapping based on this technique had been implemented by van de Meent in the Black Hole Perturbations Toolkit \cite{bhpt}.)  We have found, however, that Ferrari method solutions can be numerically unstable, including terms that have a limiting form $\epsilon/\sqrt{\epsilon^2}$ with $\epsilon$ approaching zero.  Although the {\it Mathematica} implementation can handle this ``0/0'' behavior (which has a well-behaved, finite limit) without difficulty, roundoff error leads to occasional inaccurate roots in the {\tt C}-code implementation of Ferrari's method.  We have not encountered any such difficulties using Eqs.\ (\ref{eq:r_a_gen})--(\ref{eq:r_4_gen}).  {\it Mathematica} and {\tt C}-code implementations of these solutions are provided in the supplemental material accompanying this manuscript.

\section{Comparing adiabatic inspirals}
\label{sec:compare}

We now examine adiabatic inspiral constructed by integrating the rates of change of orbit integrals ($dE/dt$, $dL_z/dt$, $dQ/dt$) to build [$E(t)$, $L_z(t)$, $Q(t)$], and by integrating the rates of change of orbit geometry ($dp/dt$, $de/dt$, $dx_I/dt$) to build [$p(t)$, $e(t)$, $x_I(t)$].  For this comparison, we use data computed for the FEW project, including data for an ongoing extension to Kerr.  We also use data that was originally developed and used in Ref.\ \cite{hughes2021}.  As such, we present results for three data sets:

\begin{enumerate}

\item {\it Equatorial Schwarzschild}: These data are used in the FEW package.  The data are computed on a grid in the ($p$, $e$) plane; there are 40 points in the $e$ direction over the domain $0 \le e \le 0.8$, evenly spaced in $e^2$ to yield denser coverage at small eccentricity.  There are 36 points in the $p$ direction, spaced according to the formula
\begin{equation}
p_j = p_{\rm min} + 4M (e^{j\Delta u} - 1)\;,\;\; 0\le j \le 35\;.
\end{equation}
We use $\Delta u = 0.035$, and set $p_{\rm min} = p_{\rm LSO} + \Delta p$.  Until recently, our frequency-domain code for solving the Teukolsky equation experienced numerical difficulty for eccentric orbits very close to the LSO.  Most of the data we have generated accordingly are on a grid whose inner edge is shifted from the LSO by $\Delta p = 0.05M$.  We have recently fixed this difficulty (which was related to resolving the Teukolsky source function when integrating orbits very close to the LSO), and have begun making data with much smaller $\Delta p$.  Because closeness to the LSO is particularly relevant for examining the singular Jacobian, we also present Schwarzschild equatorial results for $\Delta p = 10^{-4}M$.

\item {\it Equatorial Kerr, $a = 0.9M$, prograde}: These data were developed for an ongoing extension of FEW to Kerr.  The grid is identical to that used for equatorial Schwarzschild, but with $p_{\rm LSO}$ adjusted to a form appropriate for Kerr black hole orbits.  We present results only for $\Delta p = 0.05M$.

\item {\it Generic Kerr, $a = 0.7M$}: The generic grid is in an older format, covering the range $0 \le e \le 0.4$, uniformly spaced with $\Delta e = 0.1$ (a total of 5 grid points in eccentricity); uniformly spaced in inclination from retrograde to prograde, $-1 \le x_I \le 1$ with $\Delta x_I = 2/15$ (16 grid points in inclination); and with 50 grid points in $p$, uniformly spaced in
\begin{equation}
u = \frac{1}{\sqrt{p - 0.9p_{\rm LSO}}}\;.
\end{equation}
The generic grid covers $p_{\rm min} = p_{\rm LSO} + 0.02M \le p \le p_{\rm max} = p_{\rm min} + 10M$.

\end{enumerate}

At each point on these grids, we solve the frequency domain Teukolsky equation \cite{teuk73}, including enough modes to achieve flux convergence of roughly $10^{-5}$; see Ref.\ \cite{hughes2021} for detailed discussion of our procedure.  We thus have data describing ($dE/dt$, $dL_z/dt$, $dQ/dt$) at each orbit on our grids; by applying the Jacobian, we also have data describing ($dp/dt$, $de/dt$, $dx_I/dt$) at each orbit.

\begin{figure*}
\includegraphics[scale=0.4]{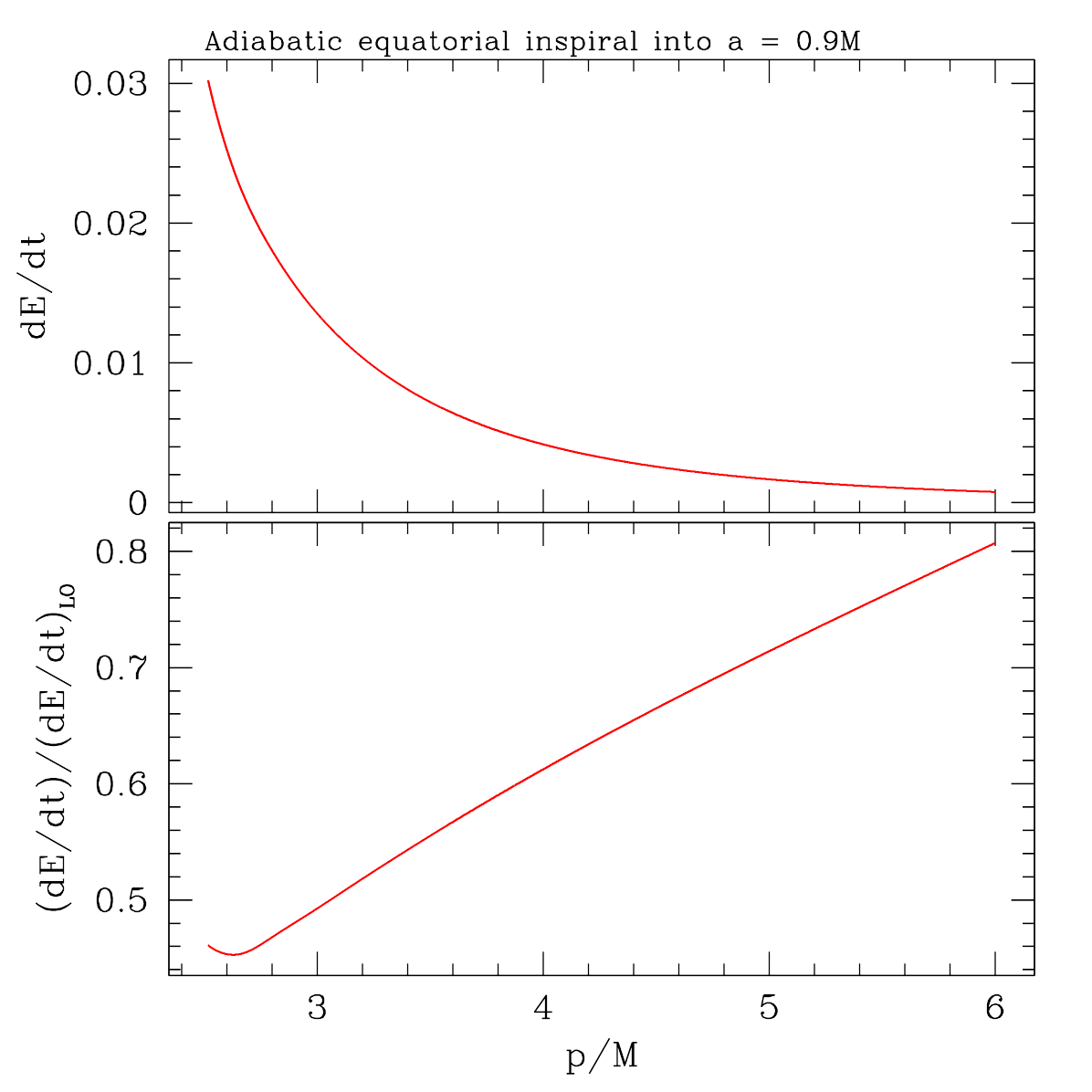}
\includegraphics[scale=0.4]{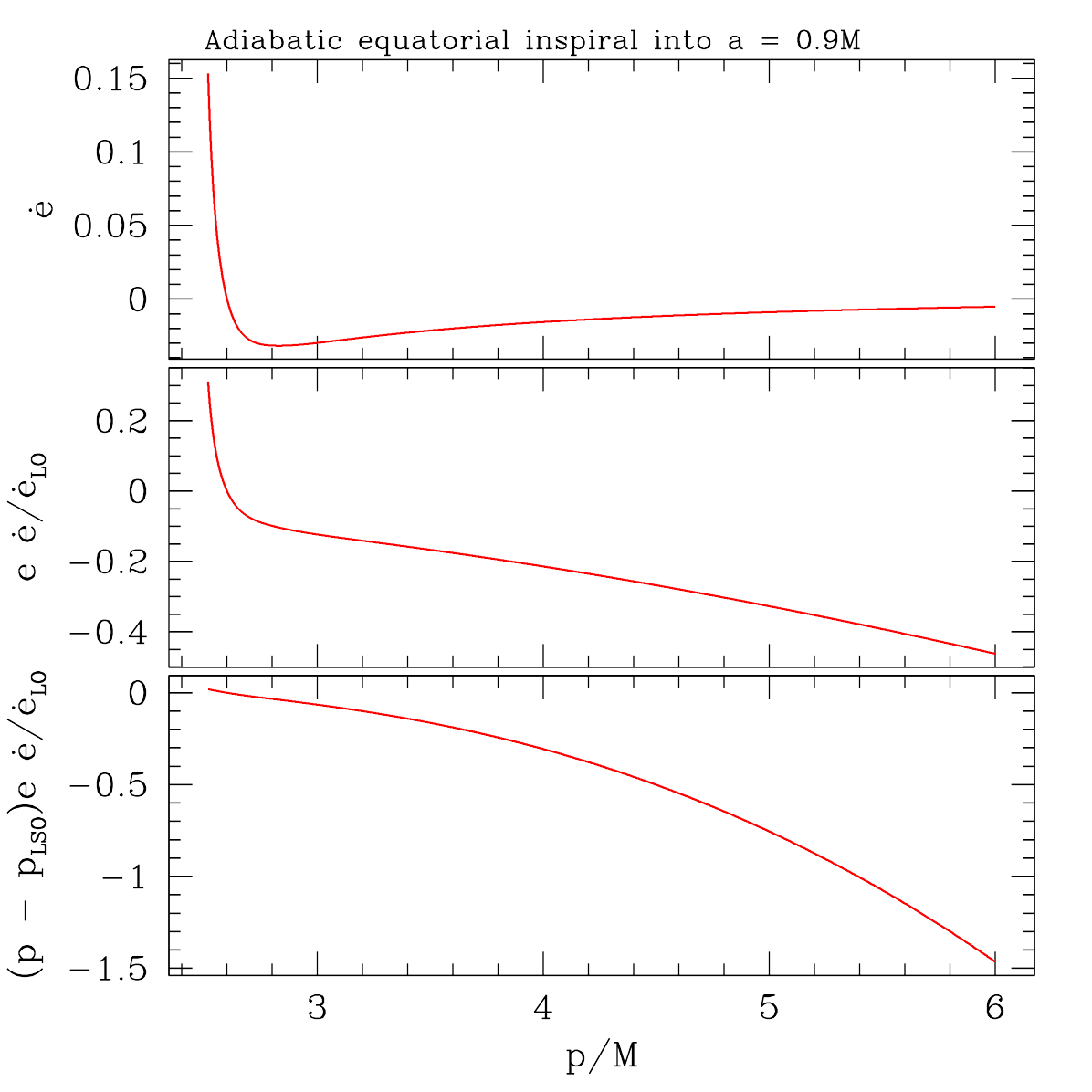}
\caption{The change of energy, $dE/dt$, and of orbital eccentricity, $\dot e \equiv de/dt$, along prograde equatorial adiabatic inspiral into a black hole with $a = 0.9M$; further details of this inspiral are shown in Fig.\ \ref{fig:a0.9_insp}.  Top left shows $dE/dt$; bottom left shows the same data normalized by the weak-field formula $(dE/dt)_{\rm LO}$ from Eq.\ (\ref{eq:dEdtLO}).  The normalized data are significantly flattened, showing much less variation than ``raw'' data.  Right-hand panels show $de/dt$: top shows ``raw'' $de/dt$ along the inspiral; middle shows these data normalized by $(1/e)(de/dt)_{\rm LO}$, from Eq.\ (\ref{eq:dedtLO}); and bottom shows the normalized data multiplied by $p - p_{\rm LSO}(e)$.  Normalizing to the weak-field form greatly reduces the variation for $p \gtrsim 2.7M$, but a large spike remains at the end of inspiral.  This spike is due to the singular Jacobian which converts $(dE/dt, dL_z/dt)$ to $(dp/dt, de/dt)$, and scales as $1/[p - p_{\rm LSO}(e)]$.  The bottom-right panel shows that the multiplicative factor significantly ameliorates the near-LSO spike.}
\label{fig:Edot_and_eccdot_insp}
\end{figure*}

To obtain data away from grid points, we use cubic spline interpolation along ($p$, $e$, $x_I$).  Because the data vary significantly over the grid, we flatten them by dividing each datum with their lowest order form.  For most data, the lowest order form is obtained by applying the quadrupole formula to orbits in Newtonian gravity \cite{peters64}.  Inclination is constant at lowest order, so next-order terms provide a low-order estimate of $dx_I/dt$ \cite{ryan96}:
\begin{widetext}
\begin{eqnarray}
\left(\frac{dE}{dt}\right)_{\rm LO} &=& -\frac{32}{5}\eta^2\left(\frac{M}{p}\right)^5(1 - e^2)^{3/2}\left(1 + \frac{73}{24}e^2 + \frac{37}{96}e^4\right)\;,
\label{eq:dEdtLO}\\
\frac{1}{x_I}\left(\frac{dL_z}{dt}\right)_{\rm LO} &=& -\frac{32}{5}\eta^2M\left(\frac{M}{p}\right)^{7/2}(1 - e^2)^{3/2}\left(1 + \frac{7}{8}e^2\right)\;,
\label{eq:dLzdtLO}\\
\frac{1}{(1 - x_I^2)}\left(\frac{dQ}{dt}\right)_{\rm LO} &=& -\frac{64}{5}\eta^3M^3\left(\frac{M}{p}\right)^3(1 - e^2)^{3/2}\left(1 + \frac{7}{8}e^2\right)\;,
\label{eq:dQdtLO}\\
\left(\frac{dp}{dt}\right)_{\rm LO} &=& -\frac{64}{5}\eta\left(\frac{M}{p}\right)^3(1 - e^2)^{1/2}\left(1 - \frac{1}{8}e^2 - \frac{7}{8}e^4\right)\;,
\label{eq:dpdtLO}\\
\frac{1}{e}\left(\frac{de}{dt}\right)_{\rm LO} &=& -\frac{304}{15}\frac{\eta}{M}\left(\frac{M}{p}\right)^4(1 - e^2)^{3/2}\left(1 + \frac{121}{304}e^2\right)\;,
\label{eq:dedtLO}\\
\frac{1}{a}\left(\frac{1}{1-x_I^2}\right)\left(\frac{dx_I}{dt}\right)_{\rm LO} &=& -\frac{244}{15}\frac{\eta}{M}\left(\frac{M}{p}\right)^{11/2}(1 - e^2)^{3/2}\left(1 + \frac{189}{61}e^2 + \frac{285}{488}e^4\right)\;.
\label{eq:dxIdtLO}
\end{eqnarray}
\end{widetext}
This procedure was introduced in the original presentation of FEW \cite{FEW1}, and has been found to be a valuable tool for developing accurate interpolants on our data grids.  Thus, for example, rather than developing spline fits to our numerical data for $dE/dt$, we develop spline fits to $(dE/dt)/(dE/dt)_{\rm LO}$.  Several of the low-order formulas that we use go to zero within the grid: $(dL_z/dt)_{\rm LO}$ vanishes at $x_I = 0$, $(dQ/dt)_{\rm LO}$ and $(dx_I/dt)_{\rm LO}$ vanish at $x_I = \pm 1$, $(dx_I/dt)_{\rm LO}$ vanishes at $a = 0$, and $(de/dt)_{\rm LO}$ vanishes at $e = 0$.  To avoid dividing by zero, in Eqs.\ (\ref{eq:dLzdtLO}), (\ref{eq:dQdtLO}), (\ref{eq:dedtLO}), and (\ref{eq:dxIdtLO}) we have factored out the contributions that remain non-zero everywhere on the grid, and flatten using only these contributions.  Thus, for example, we develop spline fits to $(dL_z/dt)/[(1/x_I)(dL_z/dt)_{\rm LO}]$.  Some examples of how these different data fields vary over an inspiral are shown in Fig.\ \ref{fig:Edot_and_eccdot_insp}, as well as how their variations can be flattened by normalizing with low-order formulas.

Note that Eqs.\ (\ref{eq:dEdtLO})--(\ref{eq:dxIdtLO}) have been converted from the forms presented in Refs.\ \cite{peters64, ryan96} by taking the small mass ratio limit ($m_1 \ll m_2$) and changing semi-major axis to semi-latus rectum using $a_{\rm axis} = p/(1 - e^2)$.  (We write $a_{\rm axis}$ for the semi-major axis to avoid confusion with the Kerr parameter.)  We have introduced the reduced mass ratio $\eta = \mu/M$, with $M = m_1 + m_2$ and $\mu = m_1m_2/M$.  Note that the result for $(dQ/dt)_{\rm LO}$ follows from the fact that at leading order $Q = |{\bf L}|^2 - L_z^2$, where ${\bf L}$ is an orbit's total angular momentum.  The result for $(dx_I/dt)_{\rm LO}$ uses the fact that our inclination angle $I$ is equivalent to the angle $\iota$ of Ref.\ \cite{ryan96} in the weak field.

In the Schwarzschild limit, it is not hard to show that the Jacobian from ($dE/dt$, $dL_z/dt$, $dQ/dt$) to ($dp/dt$, $de/dt$, $dx_I/dt$) introduces terms proportional to $1/[p - p_{\rm LSO}(e)]$, where $p_{\rm LSO}(e) = (6+2e)M$ for $a = 0$.  This singularity only affects $dp/dt$ and $de/dt$; $dx_I/dt$ is not affected.  Under the hypothesis that this scaling holds more generally, this suggests that a further useful flattening of the data is to multiply $dp/dt$ and $de/dt$ by $p - p_{\rm LSO}(e, x_I)$ before splines are constructed.  Figure \ref{fig:Edot_and_eccdot_insp} includes an example showing how this factor reduces the singular spikiness of $de/dt$ near the LSO.  We then divide by $p - p_{\rm LSO}(e, x_I)$ after spline interpolation to fully compute each off-grid datum.

We now compare inspirals computed by using the rates of change of the orbit integrals, ($dE/dt$, $dL_z/dt$, $dQ/dt$), and by using the rates of change of the orbit geometry, ($dp/dt$, $de/dt$, $dx_I/dt$).  In both cases, we use precomputed data stored on a grid parameterized in ($p$, $e$, $x_I$) as described above.  For the inspirals computed using ($dE/dt$, $dL_z/dt$, $dQ/dt$), we use Eqs.\ (\ref{eq:xIfunc}) and (\ref{eq:pe_of_rarp}) in conjunction with either Eqs.\ (\ref{eq:r_a_eq})--(\ref{eq:r_3_eq}) or Eqs.\ (\ref{eq:r_a_gen})--(\ref{eq:r_4_gen}) to determine the orbit's geometry.  In all cases, we use a simple 4th-order Runge-Kutta integrator, with step size given by
\begin{equation}
dt = dt_*\left(\frac{p}{3M}\right)^3\;.
\end{equation}
In the weak-field, circular limit, this scaling yields data with constant spacing in $p$ as inspiral proceeds; note that $(dp/dt)_{\rm LO}$ scales as $p^{-3}$.  Though this scaling does not precisely describe strong field and non-circular inspirals, we have found it nonetheless effectively compensates for the rate at which inspiral accelerates.  When we get close to the inner edge of our grid ($p - p_{\rm edge} < 0.1M$), we decrease $dt_*$ by a factor of 10 to more closely approach the edge of our dataset.  We repeat this when $p - p_{\rm edge} < 0.01M$, terminating the inspiral when $p - p_{\rm edge} < 0.001M$.  We start with $dt_* = 0.005M^2/\mu$.

\begin{figure}[ht]
\includegraphics[scale=0.4]{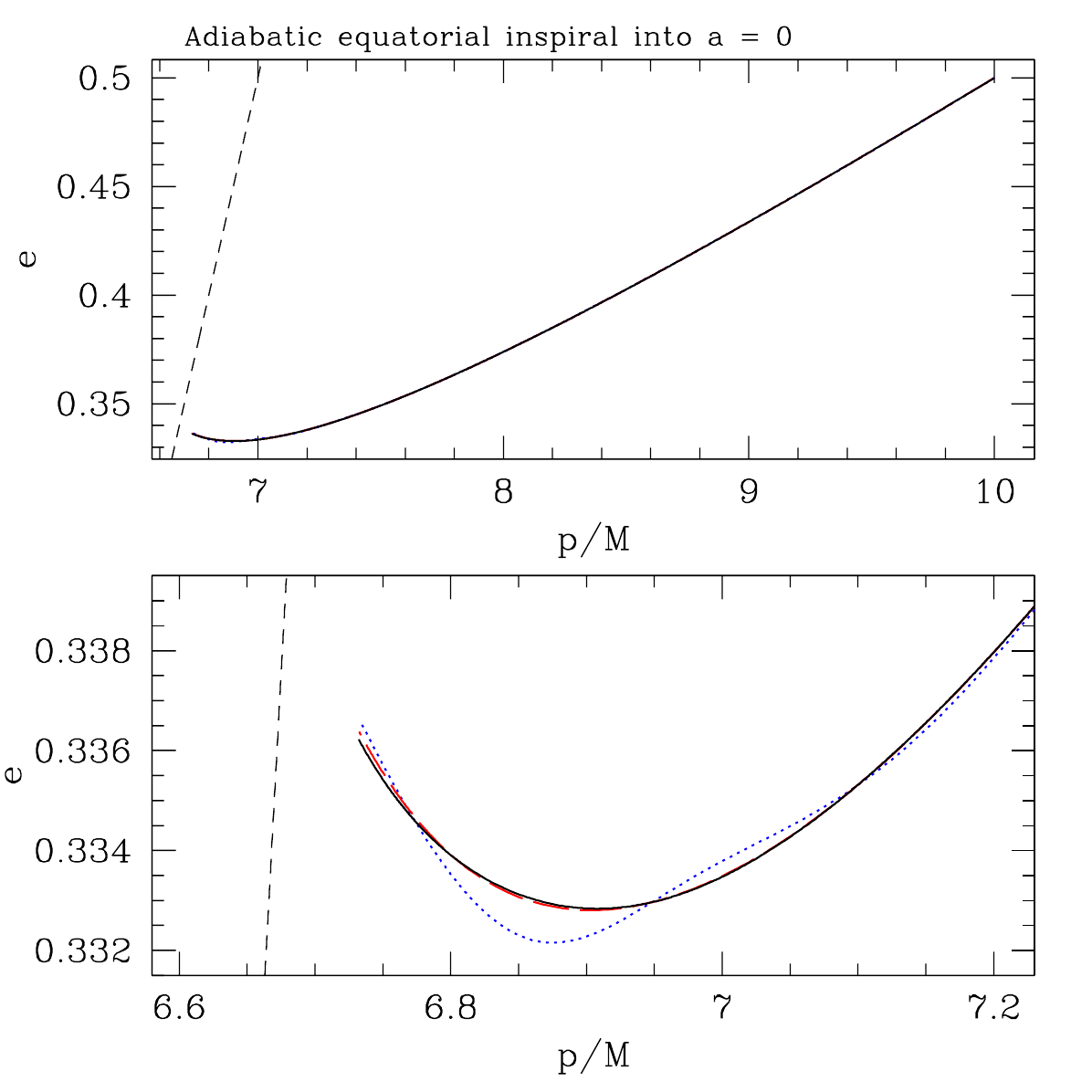}
\caption{Comparison of equatorial inspiral into a black hole with $a = 0$, computed using ($dE/dt$, $dL_z/dt$) (solid black curve), using ($dp/dt$, $de/dt$) (dotted blue curve), and using ($dp/dt$, $de/dt$) flattened by $p - p_{\rm LSO}(e)$ prior to spline interpolation (dashed red curve) to remove effects of the singular Jacobian at the LSO (black dashed line).  Top panel shows complete inspirals from $p = 10M$, $e = 0.5$ until they hit the edge of our data grid near the LSO.  The different methods are clearly distinguished in the bottom panel by zooming into the end of inspiral.  Inspiral constructed using ($dE/dt$, $dL_z/dt$) and using ($dp/dt$, $de/dt$) flattened by $p - p_{\rm LSO}(e)$ are quite similar on this plot.  However, inspiral using ($dp/dt$, $de/dt$) without correcting for the singular Jacobian shows significant differences.  The variations seen in the dotted blue curve are numerical artifacts of fitting a cubic spline to such spiky data.}
\label{fig:a0.0_insp}
\end{figure}

We begin with equatorial inspiral, for which $Q = 0$ and $x_I = 1$ (we focus on prograde cases in this brief study).  Figure \ref{fig:a0.0_insp} looks at inspiral into a Schwarzschild black hole.  We show three inspirals: inspiral $I$ is computed by integrating the rates of change of orbit integrals, ($dE/dt$, $dL_z/dt$); inspiral $G$ is computed by integrating the rates of change of orbit geometry, ($dp/dt$, $de/dt$); and inspiral $G_F$ is computed by integrating the rates of change ($dp/dt$, $de/dt$), but further flattening all grid data by $p - p_{\rm LSO}(e)$.  All inspirals start at $p = 10M$, $e = 0.5$, and proceed until they reach the inner edge of the data grid (which, as discussed above, is shifted away from the LSO by $\Delta p = 0.05M$).  Inspiral $I$ is shown as a solid black curve, inspiral $G$ is a dotted blue curve, and inspiral $G_F$ is a dashed red curve.

The top panel shows the full span of all three inspirals.  The three curves can barely be distinguished at this scale; these three methods of treating the data yield inspirals that are nearly identical, at least for $p \gtrsim 7.5M$ or so.  The bottom panel zooms in on the end of inspiral, showing significant differences between inspiral $G$ and the others, and smaller differences between inspirals $I$ and $G_F$.  The variations between $G$ and the other inspirals are numerical artifacts that arise from attempting to fit such spiky data with a cubic spline.  Interestingly, inspirals $I$ and $G_F$ are quite similar.  Removing the singularity with the factor $p - p_{\rm LSO}(e)$ yields an inspiral using ($dp/dt$, $de/dt$) that is nearly identical to that found using ($dE/dt$, $dL_z/dt$).

\begin{figure}[ht]
\includegraphics[scale=0.4]{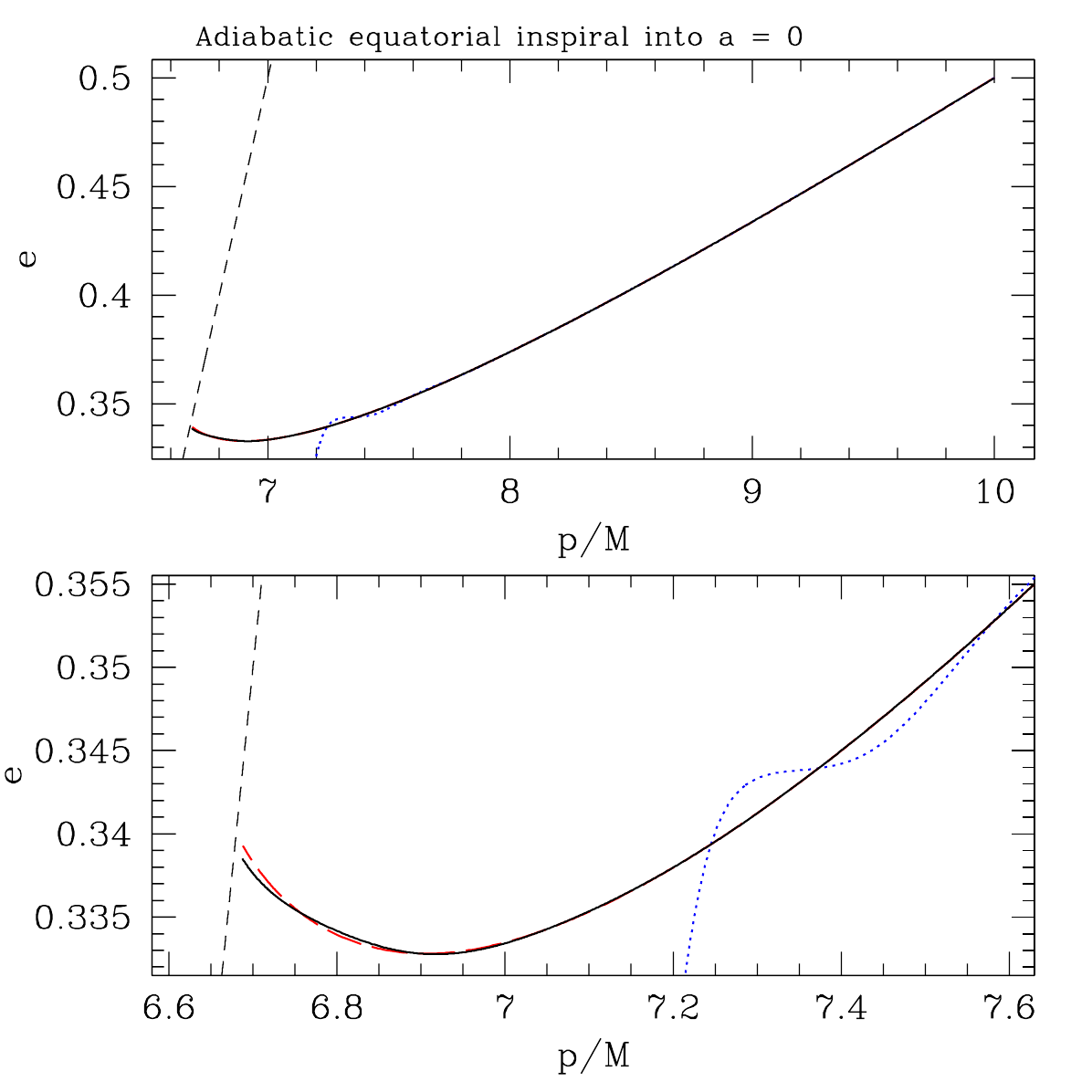}
\caption{Same as Fig.\ \ref{fig:a0.0_insp}, but using a grid that comes much closer to the LSO ($\Delta p = 10^{-4}M$ rather than $\Delta p = 0.05M$).  Inspiral constructed using ($dp/dt$, $de/dt$) without accounting for the $p - p_{\rm LSO}$ singularity is totally pathological here: the dotted blue curve showing this inspiral noticeably deviates even in the large scale top panel.  Lower panel, zooming on the end of inspiral, shows more detail of this evolution.  Evolution using ($dE/dt$, $dL_z/dt$) (solid black curve) and using ($dp/dt$, $de/dt$) flattened by $p - p_{\rm LSO}(e)$ (dashed red curve) are again similar, though with larger differences than seen in Fig.\ \ref{fig:a0.0_insp}.}
\label{fig:a0.0_insp_v2}
\end{figure}

Figure \ref{fig:a0.0_insp_v2} re-examines Schwarzschild inspiral, but now using data that come closer to the LSO ($\Delta p = 10^{-4}M$).  Aside from the deeper span of data, the inspirals are identical to those shown in Fig.\ \ref{fig:a0.0_insp}.  Inspirals $I$ and $G_F$ are not significantly affected by the new data, though the difference between these two inspirals is slightly larger than the difference seen when $\Delta p = 0.05M$.  By contrast, the numerical artifacts which affect inspiral $G$ are so large that the inspiral's behavior is quite pathological: it begins oscillating at $p \simeq 7.5M$, $e \simeq 0.345$, and deviates completely from the other inspirals.  Cubic spline fits to data processed by the singular Jacobian are extremely ill behaved for data that comes so close to the LSO.

\begin{figure}[h]
\includegraphics[scale=0.4]{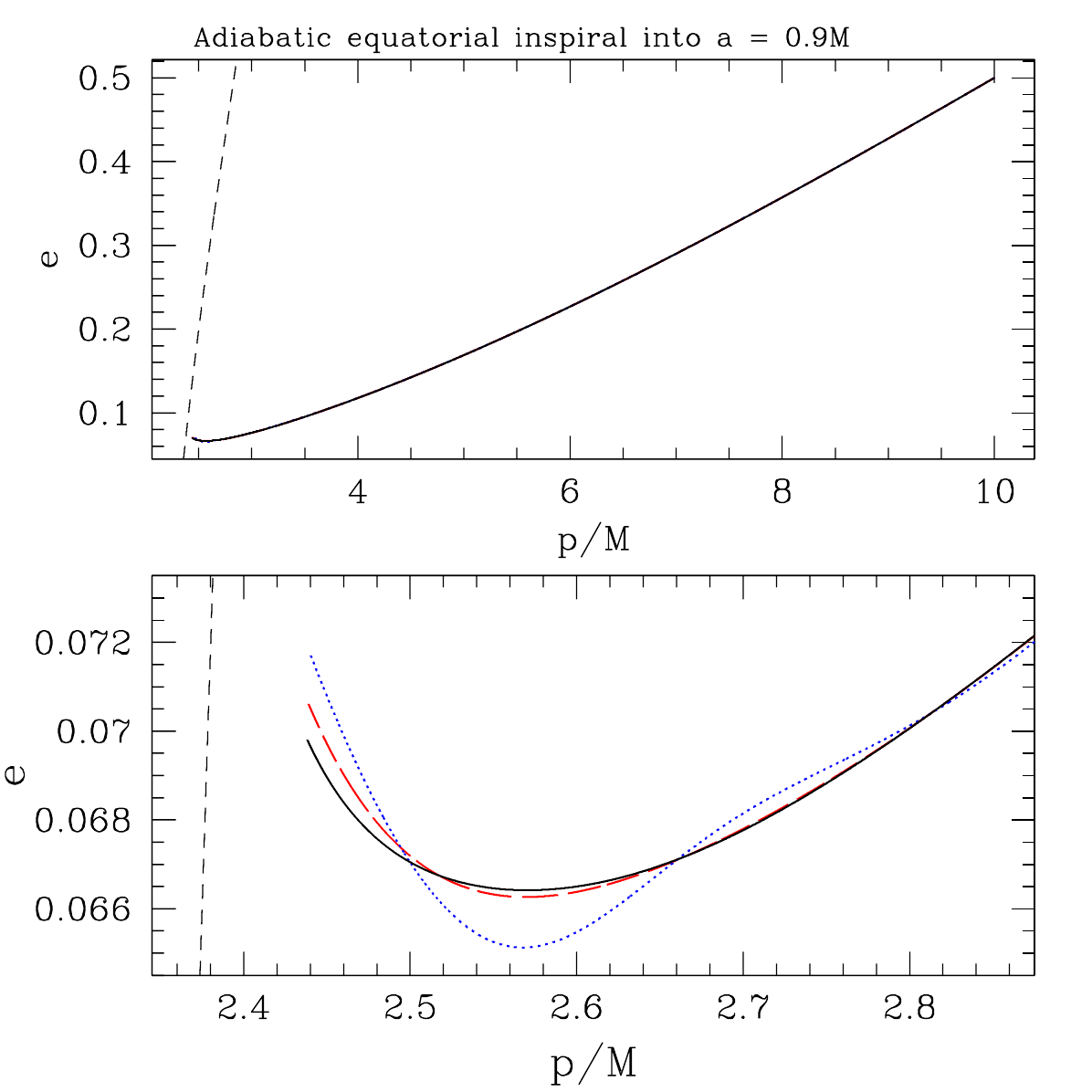}
\caption{Comparison of equatorial inspiral into a black hole with $a = 0.9M$; aside from the different black hole spin, the parameters are the same as those used in Fig.\ \ref{fig:a0.0_insp}.  We again see that these inspirals are similar until they come close to the LSO: inspiral constructed using ($dE/dt$, $dL_z/dt$) (solid black curve) and using ($dp/dt$, $de/dt$) with data flattened by $p - p_{\rm LSO}(e)$ prior to spline interpolation (dashed red curve) differ significantly from data computed using ($dp/dt$, $de/dt$) without flattening (dotted blue curve).  However, the difference between the ($dE/dt$, $dL_z/dt$) inspiral and the flattened ($dp/dt$, $de/dt$) inspiral is somewhat larger than we saw in the Schwarzschild inspiral}
\label{fig:a0.9_insp}
\end{figure}

Figure \ref{fig:a0.9_insp} likewise examines equatorial inspiral from $p = 10M$, $e = 0.5$, but now into a Kerr black hole with $a = 0.9M$.  The pattern we see is largely the same as what we saw in the Schwarzschild cases, but because the LSO is at smaller $p$, inspiral proceeds much deeper into the strong field.  In particular, we see strong oscillations in $G$ in the deep strong field.  Inspirals $I$ and $G_F$ are similar to each other, though one can see that inspiral $G_F$ shows a small amplitude oscillation about the trajectory defined by inspiral $I$.  These oscillations are nearly identical in period to the oscillations seen in inspiral $G$, suggesting that the flattening obtained by using $p - p_{\rm LSO}(e)$ does not fully correct for the singular Jacobian.

\begin{figure}[h]
\includegraphics[scale=0.4]{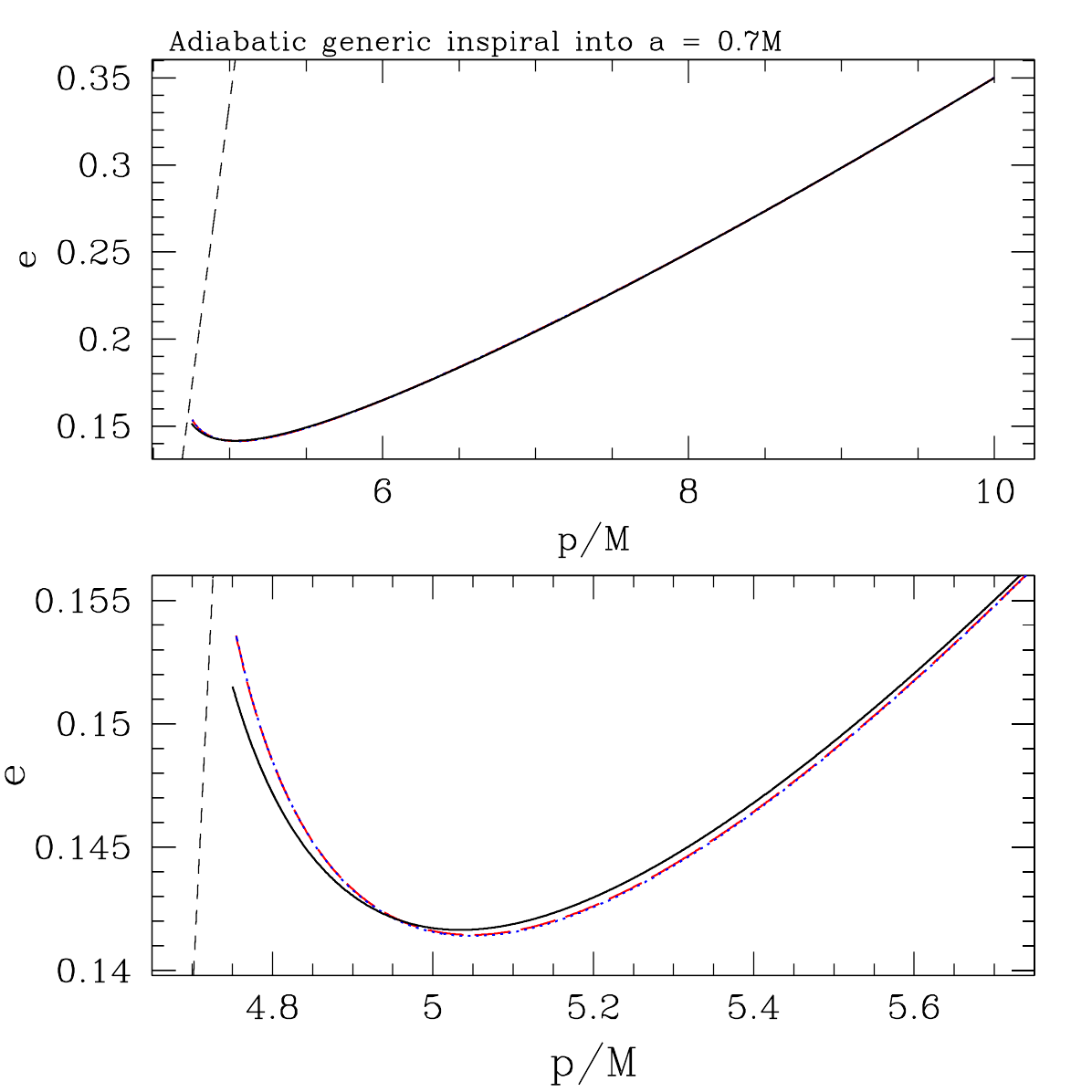}
\caption{Comparison of generic inspiral into a black hole with $a = 0.7M$.  The color scheme is similar to previous figures: solid black shows an inspiral computed using ($dE/dt$, $dL_z/dt$, $dQ/dt$); dotted blue shows an inspiral computed using ($dp/dt$, $de/dt$, $dx_I/dt$); and dashed red shows an inspiral that uses ($dp/dt$, $de/dt$, $dx_I/dt$), with ($dp/dt$, $de/dt$) flattened by $p - p_{\rm LSO}(e, x_I)$.  These three inspirals are very similar, though deviations can be discerned near the end.  Interestingly, the inspiral computed using ($dp/dt$, $de/dt$, $dx_I/dt$) barely differs from the inspiral computed using these data flattened by $p - p_{\rm LSO}(e, x_I)$; at least at this inclination, the effect of singularity flattening does not appear to be significant.  It is worth noting that both of the inspirals computed using the rates of change of orbit geometry oscillate with small amplitude around the inspiral computed using the rate of change of orbit integrals.  Note also that inclination barely changes during inspiral, decreasing from $x_I = 0.5$ to $x_I = 0.49$ at the end of inspiral.  The three tracks along $x_I$ are so similar that we do not show them.}
\label{fig:a0.7_insp}
\end{figure}

We conclude our examination of inspirals with a generic example: inspiral into a black hole with $a = 0.7M$.  As shown in Fig.\ \ref{fig:a0.7_insp}, these inspirals begin at $p = 10M$, $e = 0.35$, $x_I = 0.5$, and reach the edge of our data set near the LSO at $p \simeq 4.75M$, $e \simeq 0.152$, $x_I \simeq 0.49$.  Interestingly, we see that inspirals $G$ and $G_F$ barely differ --- this flattening of the data does not have much impact on the results.  Both cases differ slightly from $I$.  Careful inspection shows that both inspirals $G$ and $G_F$ oscillate around inspiral $I$, likely a numerical artifact arising from cubic spline interpolation applied to the singular Jacobian.

In all cases that we examine, inspirals generated using ($dE/dt$, $dL_z/dt$, $dQ/dt$) are free of numerical artifacts and well behaved.  Inspirals generated using ($dp/dt$, $de/dt$, $dx_I/dt$) are often affected by numerical artifacts arising from the singular behavior of the Jacobian between the two parameterizations near the LSO.  The effects of these artifacts can be tamed somewhat by multiplying data by the factor $p - p_{\rm LSO}$ prior to developing spline interpolants, but even data flattened in this way show signs of oscillations which are characteristic of spline interpolation applied to spiky data.

\section{Conclusions}
\label{sec:conclusions}

Recent progress has significantly advanced our ability to compute precise gravitational waveforms in the small mass ratio limit.  As a matter of principle, the construction of adiabatic waveforms is essentially fully understood.  Such waveforms serve as a foundation for computing post-adiabatic effects, and for examining issues in science and data analysis with future low-frequency gravitational wave measurements.  As such, it is of great importance to understand systematic errors and limitations which affect these waveform models.

Though understood in principle, many details must come together to expedite developing these waveform models in practice.  A major systematic limitation in computing adiabatic waveforms is our ability to accurately evolve a system from orbit to orbit due to the backreaction of gravitational waves.  Leading order backreaction computes the rates of change of orbit integrals, ($dE/dt$, $dL_z/dt$, $dQ/dt$).  If these quantities were known everywhere, it would be a simple matter to integrate them up to compute the trajectory through orbit integral space, [$E(t)$, $L_z(t)$, $Q(t)$].  In fact, the parameter space is typically discretely sampled, so that these data are only known on some finite grid.  These grids tend to be laid out in coordinates describing the orbit geometry, ($p$, $e$, $x_I$).  One thus needs accurate methods for converting between the parameterizations, and accurate techniques for interpolating data off the grid points.

Accurate formulas expressing the orbit integrals as functions of orbit geometry --- $E(p, e, x_I)$, etc.\ --- have been known for quite some time \cite{schmidt, fujitahikida, vandemeent}.  Because of this, many adiabatic inspirals to date have been made by converting ($dE/dt$, $dL_z/dt$, $dQ/dt$) to ($dp/dt$, $de/dt$, $dx_I/dt$) using the Jacobian between these sets of rates of change (cf.\ Appendix B of Ref.\ \cite{hughes2021}).  This makes it possible to find the trajectory in orbit geometry space, [$p(t)$, $e(t)$, $x_I(t)$].  Unfortunately, this Jacobian is singular near the LSO.  Though its singularity can be accounted for at least in part, using this Jacobian is very likely to introduce systematic errors into inspirals, and thus into waveform models.  These errors are particularly severe if moderately stiff methods like cubic spline interpolation are used to compute quantities away from on-grid data points --- attempting to fit a spline to the near-LSO singular spike introduces oscillations which can persist even after applying tricks which remove most of the singular behavior.

In this paper, we have developed ``inverse'' mappings of the orbit parameterizations, $p(E, L_z, Q)$ and likewise for $e$ and $x_I$.  The mappings are particularly simple for equatorial orbits, but can be implemented quite efficiently for all bound Kerr orbits.  Using them, we can easily integrate up to find [$E(t)$, $L_z(t)$, $Q(t)$] and then convert to [$p(t)$, $e(t)$, $x_I(t)$] in a way that avoids using the singular Jacobian and thus avoids introducing systematic error due to its singularity.  Comparing a sample of inspirals computed using data developed for the FEW project, we find that inspirals constructed by integrating ($dE/dt$, $dL_z/dt$, $dQ/dt$) do not show behavior associated with artifacts of the near-LSO singularity.  This method of constructing inspirals appears to be quite accurate and robust, significantly expanding the ways one can construct adiabatic inspirals using backreaction data.  To facilitate their use by the community, we provide code implementing these formulas in this paper's supplemental material.

\section*{Acknowledgments}

We thank the participants in the Fast EMRI Waveform project for valuable discussions, particularly Christian Chapman-Bird, Lorenzo Speri, and Jonathan Thompson for discussions related to the foundations of this analysis, and to Josh Mathews and Soichiro Isoyama for very useful discussions about the roots of quartic polynomials.  We are especially grateful to Isoyama for alerting us to Ref.\ \cite{wolters}, and for very helpful discussion about the relative merits of different quartic root finders.  We also thank Maarten van de Meent for alerting us to to the {\it Mathematica} functionality for finding inverse mappings which he previously incorporated into the Black Hole Perturbations Toolkit, and to this paper's referee for helpful feedback and comments.  This work was supported by NSF Grant PHY-2110384 and by MIT's Carl G.\ and Shirley Sontheimer Research Fund.  Many of the algebraic manipulations presented here were done using {\it Mathematica}.

\end{document}